\documentclass[12pt,a4paper]{article}
\pdfoutput=1

\usepackage[utf8x]{inputenc}
\usepackage[T1]{fontenc}
\usepackage{authblk}

\usepackage{amsmath}
\usepackage{amssymb}
\allowdisplaybreaks[4]

\usepackage{chngcntr} 

\usepackage{natbib}
\setcitestyle{authoryear,open={(},close={)}} 

\usepackage{url}

\usepackage{subfigure}

\usepackage{caption} 
\usepackage{xcolor}

\usepackage{algorithm}
\usepackage{algorithmic}

\usepackage[pdftex]{graphicx} 
\usepackage{calc} 
\usepackage{enumitem} 

\usepackage[a4paper, lmargin=0.1\paperwidth, rmargin=0.1\paperwidth, tmargin=0.1\paperheight, bmargin=0.1\paperheight]{geometry} 

\usepackage[all]{nowidow} 
\usepackage[protrusion=true,expansion=true]{microtype} 

\usepackage{multirow}

\usepackage[title]{appendix}

\makeatletter
\newenvironment{breakablealgorithm}
  {
   \begin{center}
     \refstepcounter{algorithm}
     \hrule height.8pt depth0pt \kern2pt
     \renewcommand{\caption}[2][\relax]{
       {\raggedright\textbf{\ALG@name~\thealgorithm} ##2\par}%
       \ifx\relax##1\relax 
         \addcontentsline{loa}{algorithm}{\protect\numberline{\thealgorithm}##2}%
       \else 
         \addcontentsline{loa}{algorithm}{\protect\numberline{\thealgorithm}##1}%
       \fi
       \kern2pt\hrule\kern2pt
     }
  }{
     \kern2pt\hrule\relax
   \end{center}
  }
\makeatother

\usepackage{color}

\newcommand{\jj}{\color{black}}

\title{A Multi-Arm Two-Stage (MATS) Design for Proof-of-Concept and Dose Optimization in Early-Phase Oncology Trials}
\author[1]{Zhenghao Jiang}
\author[2]{Gu Mi}
\author[2]{Ji Lin}
\author[3]{Christelle Lorenzato}
\author[4]{Yuan Ji*}
\affil[1]{Department of Statistics, University of Chicago, 5747 South Ellis Avenue, Chicago, IL 60637}
\affil[2]{Biostatistics and Programming, Sanofi, 450 Water Street, Cambridge, MA 02141}
\affil[3]{Biostatistics and Programming, Sanofi R\&D, Vitry-sur-Seine, France}
\affil[4]{Department of Public Health Science, University of Chicago, 5841 South Maryland Avenue, Chicago, IL 60637}
\affil[*]{Correspondence to yji@bsd.uchicago.edu}
\date{}

\begin{document}

\maketitle

\section*{Abstract}
The Project Optimus initiative by the FDA's Oncology Center of Excellence is widely viewed as a groundbreaking effort to change the \textit{status quo} of conventional dose-finding strategies in oncology. Unlike in other therapeutic areas where multiple doses are evaluated thoroughly in dose ranging studies, early-phase oncology dose-finding studies are characterized by the practice of identifying a single dose, such as the  maximum tolerated dose (MTD) or the recommended phase 2 dose (RP2D). 
Following the spirit of Project Optimus, we propose an Multi-Arm Two-Stage (MATS) design for proof-of-concept (PoC) and dose optimization that allows the evaluation of two selected doses from a dose-escalation trial. The design assess the higher dose first across multiple indications in the first stage, and adaptively enters the second stage for an indication if the higher dose exhibits promising anti-tumor activities. In the second stage, a randomized comparison between the higher and lower doses is conducted to achieve proof-of-concept (PoC) and dose optimization. A Bayesian hierarchical model governs the statistical inference and decision making by borrowing information across doses, indications, and stages. 
Our simulation studies show that the proposed MATS design yield desirable performance. An R Shiny application has been developed and made available at \url{https://matsdesign.shinyapps.io/mats/}. 

\section*{Keywords} Bayesian Hierarchical Model; Oncology; Project Optimus; Proof-of-Concept.

\section{Introduction}
The past decade has witnessed exciting therapeutic advances in the oncology treatment landscape highlighted by the development of molecularly targeted agents and immunotherapies based on more effective and precise mechanisms of actions (MoAs) than conventional cytotoxic chemotherapies. While there has been a plethora of novel  dose-escalation designs that accommodate various MoAs for better characterization of toxicity or toxicity/efficacy profiles, most of them are still in the pursuit of a single maximum tolerated dose (MTD). Typically, MTD becomes the \textit{de facto} recommended phase 2 dose (RP2D) for further development. In contrast, in non-oncology  therapeutic areas multiple doses are evaluated thoroughly in large dose-ranging studies. The reason why oncology traditionally resorts to MTD as RP2D is often based on the rather limited treatment options for cancer patients and the urgency of advancing novel agents that aim to address highly unmet medical needs. Another reason is that for most cytotoxic oncology drugs, MTD is theoretically the optimal dose since it provides maximum efficacy among the doses that can be tolerated based on toxicity. 

The FDA's Project Optimus \citep{fda2022projectoptimus}, initiated in 2021 by the Oncology Center of Excellence, is a historic step in challenging the \textit{status quo} of conventional dosing-finding in oncology. Dose optimization, a central topic in the Project Optimus, was brought up at the ``\textit{Hot Topics in Oncology Regulation}'' session at the American Association for Cancer Research (AACR) Annual Meeting in April 2021. As of the writing of this article in early 2023, almost two years had passed, during which time we have seen quite a few notable papers and regulatory examples on this topic, such as a \textit{New England Journal of Medicine Perspective} by \citet{shah2021drug} emphasizing ``When Less is More'', a \textit{Friends of Cancer Research} white paper by \citet{blumenthal2021optimizing} on changes of external expectations on oncology dose optimization, and a \textit{Journal of Clinical Oncology Speical Series: Statistics in Oncology} article by \citet{fourie2022improving} highlighting dose optimization strategies with recent case examples. Among many initiatives in the Project Optimus is the recommendation of evaluating multiple doses in a randomized setting to better understand the dose/exposure-activity/toxicity relationships and identify the dose that maximizes benefit-risk ratio before submission. Novel and practical statistical designs are being proposed that not only follow the spirit of Project Optimus but also aim to be ``strategically fit'' based on each project's unique features and development stages. However, the majority of proposed methods still evaluate all dose levels in a sequential manner with the mindset of ``the higher the better'' provided that the safety profile is acceptable. Simultaneous evaluation of multiple candidate doses using sizeable, randomized cohorts has been extremely rare in oncology in the past, but that theme is expected to become a ``routine'' rather than an ``exception'' after the debut of the Project Optimus. As part of early-phase development, randomized  comparison of multiple doses in dose optimization does not necessarily need to be powered for statistical errors. However, it remains critical to address key statistical considerations such as justifying sample sizes to yield satisfactory and robust decision-making for dose selection, especially in the presence of multiple indications that demand more efficient designs at early stage of drug development. 

In response to Project Optimus, by harmonizing regulatory expectations and feedback with practical considerations from an industrial sponsor's perspective, we propose an adaptive design called MATS (Multi-Arm Two-Stage) that aim to achieve dual objectives of both proof-of-concept (PoC) and dose optimization (DO) within a single study under a Bayesian hierarchical model. We assume dose-escalation  has been conducted as part of a first-in-human phase I trial and two doses have been selected for further expansion. For example, the two doses could be the MTD or (MTD-1) dose, or two doses that are both lower than the MTD.  The objectives of the expansion are to establish 1) PoC, i.e., whether the two doses exhibit anti-tumor activities in multiple indications, and 2) DO, i.e., which of the two doses is optimal in terms of efficacy. The MATS design allows for simultaneous evaluation of multiple indications in the first stage for preliminary activity (a typical situation for the expansion stage in oncology), followed by randomizing two doses for dose optimization only within the promising indication(s) in the second stage. Extensive simulations reveal that the MATS design is capable of identifying the optimal dose and correct PoC with reasonably high probabilities by taking advantage of potential information-borrowing across cohorts, across stages and across dose levels, where appropriate. A potential benefit of MATS is the savings on the sample size since in the first stage MATS only expands with a single dose (instead of two or more) across multiple indications and only indications with promising anti-tumor activities are selected for the second stage dose comparison. 

The remainder of this article is organized as follows. In Section \ref{sec:design}, we first provide an overview of the MATS design, including the underlying Bayesian hierarchical model, the proposed design, and the determination of a key tuning parameter. Scenarios for the comprehensive simulation studies are described in Section \ref{sec:sim_setup}, followed by simulation results in Section \ref{sec:sim_results}. We conclude the article in Section \ref{sec:discussion} about utilizing MATS as part of the dose optimization data package, how MATS design is consistent with the recent FDA Guidance for dose optimization \citep{fda2023optimizing},  potential extensions of MATS according to recent regulatory feedback, and a few practical considerations from an industrial sponsor's perspective.

\section{The MATS Design}\label{sec:design}

\subsection{Probability Model}
The proposed MATS design evaluates $I=2$ different dose levels of a new investigational drug in $J$ different indications at $K=2$ stages. Let $(i,j,k)$ denote the arm for dose level $i$, indication $j$ and stage $k$, $i=1,\dots ,I$, $j=1,\dots,J$, $k=1,\dots,K$. 
Denote the higher dose by ``DL-H'' corresponding to $i=1$ and a lower dose by ``DL-L'' corresponding to $i=2$.  The two doses are assumed to have been selected from a preceding dose-escalation stage in a phase I trial that evaluates multiple ascending doses based on dose-limiting toxicity (DLT). For example, the DL-H and DL-L doses may be the MTD and (MTD-1). The MATS design can then be used to guide the next stages of the trial, focusing on expansion and dose optimization. 

During Stage 1, MATS  enrolls participants at the DL-H dose across all indications. Suppose $n_{1j1}$ participants have been treated in arm $(1,j,1)$, and $y_{1j1}$ of them are showing preliminary anti-tumor activity (e.g., complete or partial responders per RECIST 1.1 for solid tumors). Let $p_{ij}$ denote the true and unknown response rate for dose $i$ and indication $j$.
We assume $y_{1j1}$ follows a binomial distribution, i.e., 

$$y_{1j1}|n_{1j1},p_{1j}\sim \text{Bin}(n_{1j1},p_{1j}).$$ 

Let $D_{j1}\in \{0,1\}$ denote the decision made at the end of Stage 1 for indication $j$, where $D_{j1}=1$ means ``go to Stage 2'' and indication $j$ is considered ``promising'', while $D_{j1}=0$ means ``do not go to Stage 2 and stop for futility'' so that no further enrollment for indication $j$ will be made. 

During Stage 2, we randomize participants to both dose levels DL-H and DL-L if $D_{j1}=1$ (i.e., only for those ``promising'' indications identified at Stage 1). Suppose $n_{ij2}$ participants have been treated in arm $(i,j,2)$, and $y_{ij2}$ of them are responders. 
We assume $y_{ij2}$ follows a binomial distribution given that  $D_{j1}=1$, i.e., Stage 1 decision is a go. Then
$$y_{ij2} |n_{ij2},p_{ij}, D_{j1}=1 \sim \text{Bin}(n_{ij2}, p_{ij}).$$

At this stage, we want to test the PoC and DO  based on two sets of hypotheses. For PoC, we consider the following null and alternative hypotheses, 
\begin{equation*}
    H_{0,ij}: p_{ij}\leq p_{0j} \quad\text{versus}\quad H_{1,ij}: p_{ij} > p_{0j},
\end{equation*}
where $p_{0j}$ is the reference response rate for indication $j$ that is elicited from clinicians and/or historical data. For DO, we  examine whether the difference in efficacy between the two dose levels for indication $j$ is sufficiently large by the following hypothesese, 
\begin{equation*}
    H_{0,j}: p_{1j}-p_{2j}\leq \epsilon \quad\text{versus}\quad H_{1,j}: p_{1j}-p_{2j}> \epsilon.
\end{equation*}
where $\epsilon$ is a predetermined threshold reflecting the desirable difference in efficacy between the two dose levels. We further assume that the efficacy rate of each dose is monotonically non-decreasing with the dose level. Therefore, $p_{1j} \ge p_{2j}$. This assumption is consistent across a wide variety of drugs such as cytotoxic agents, antibody drug conjugates, molecularly targeted agents and immunotherapies, etc., although their therapeutic index may be relatively narrower or wider due to different MoAs. When this assumption is not true, the MATS design may not be appropriate. 

Let $\theta_{ij}=\text{logit}(p_{ij})$ denote the log-odds of the response rate for $i=0,1,2$, where $\text{logit}(x)=\log[x/(1-x)]$.  We define 
\begin{equation*}
    \eta_j=\theta_{1j}-\theta_{0j} \quad\text{and}\quad \gamma_j=\theta_{1j}-\theta_{2j}
\end{equation*}
to represent the difference in the response rate (at the logit scale) between the DL-H dose and reference rate and between the two doses DL-H and DL-L, respectively. Then by definition, $(\eta_j - \gamma_j)$ quantifies the difference between the DL-L dose and the reference. 
We model $\eta_j$ and $\gamma_j$ via shrinkage priors given by
\begin{equation*}
    \eta_j|\eta_0,\sigma^2_{\eta} \sim \text{N}(\eta_0,\sigma^2_{\eta}) \quad\text{and}\quad \gamma_j|\gamma_0,\sigma^2_{\gamma} \sim \text{LogNormal}(\gamma_0,\sigma^2_{\gamma}),
\end{equation*}
which are independent across indication $j$. To induce shrinkage, we assume the following hyper-priors for parameters $\eta_0$, $\sigma^2_{\eta}$, $\gamma_0$ and $\sigma^2_{\gamma}$, i.e., 
\begin{equation*}
\begin{aligned}
    \eta_0|\mu_{\eta_0},\sigma^2_{\eta_0} \sim \text{N}(\mu_{\eta_0},\sigma^2_{\eta_0}) &\quad\text{and}\quad \sigma^2_{\eta}|\alpha_{\eta},\beta_{\eta} \sim \text{Inv-Gamma}(\alpha_{\eta},\beta_{\eta})\\
    \gamma_0|\mu_{\gamma_0},\sigma^2_{\gamma_0} \sim \text{N}(\mu_{\gamma_0},\sigma^2_{\gamma_0}) &\quad\text{and}\quad \sigma^2_{\gamma}|\alpha_{\gamma},\beta_{\gamma} \sim \text{Inv-Gamma}(\alpha_{\gamma},\beta_{\gamma})
\end{aligned}
\end{equation*}

At the end of the trial, we make a final decision $D_{j2}\in \{0,1,2\}$, where  $D_{j2}=1$ means selecting the higher dose DL-H as the optimal dose,  $D_{j2}=2$ means selecting the lower dose DL-L as the optimal dose, and $D_{j2}=0$ indicates that neither dose is selected, i.e., no sufficient efficacy signal is observed for indication $j$ at either dose. 

The entire hierarchical model of the MATS design is summarized as below: 

\begin{equation}
\begin{aligned}
    &\text{Stage 1 likelihood:} &y_{1j1}|n_{1j1},p_{1j}\sim \text{Bin}(n_{1j1},p_{1j}) \\
    &\text{Stage 2 likelihood:} &y_{ij2}|n_{ij2},p_{ij},D_{j1}=1 \sim \text{Bin}(n_{ij2}\cdot D_{j1},p_{ij}) \\
    &\text{Transformation:} &\theta_{0j}=\text{logit}(p_{0j}), \theta_{ij}=\text{logit}(p_{ij}) \\
    &\text{Prior for }\theta_{ij}: &\theta_{1j}=\theta_{0j}+\eta_j \text{ and } \theta_{2j}=\theta_{1j}-\gamma_j \\
    &\text{Indication-specific treatment effects:} &\eta_j|\eta_0,\sigma^2_{\eta} \sim \text{N}(\eta_0,\sigma^2_{\eta}) \\
    &&\gamma_j|\gamma_0,\sigma^2_{\gamma} \sim \text{LogNormal}(\gamma_0,\sigma^2_{\gamma}) \\
    &\text{Hyper-priors:} &\eta_0|\mu_{\eta_0},\sigma^2_{\eta_0} \sim \text{N}(\mu_{\eta_0},\sigma^2_{\eta_0}) \\
    &&\sigma^2_{\eta}|\alpha_{\eta},\beta_{\eta} \sim \text{Inv-Gamma}(\alpha_{\eta},\beta_{\eta}) \\
    &&\gamma_0|\mu_{\gamma_0},\sigma^2_{\gamma_0} \sim \text{N}(\mu_{\gamma_0},\sigma^2_{\gamma_0}) \\
    &&\sigma^2_{\gamma}|\alpha_{\gamma},\beta_{\gamma} \sim \text{Inv-Gamma}(\alpha_{\gamma},\beta_{\gamma})
\end{aligned} \label{eq:model}
\end{equation}

Posterior inference of the model is conducted through the Markov chain Monte Carlo (MCMC) simulations. Specifically, MCMC samples of each parameter are obtained with which posterior probabilities and decisions in subsequent sections are computed. 

\begin{figure}[htbp]
    \centering
    \includegraphics[width=0.8\textwidth]{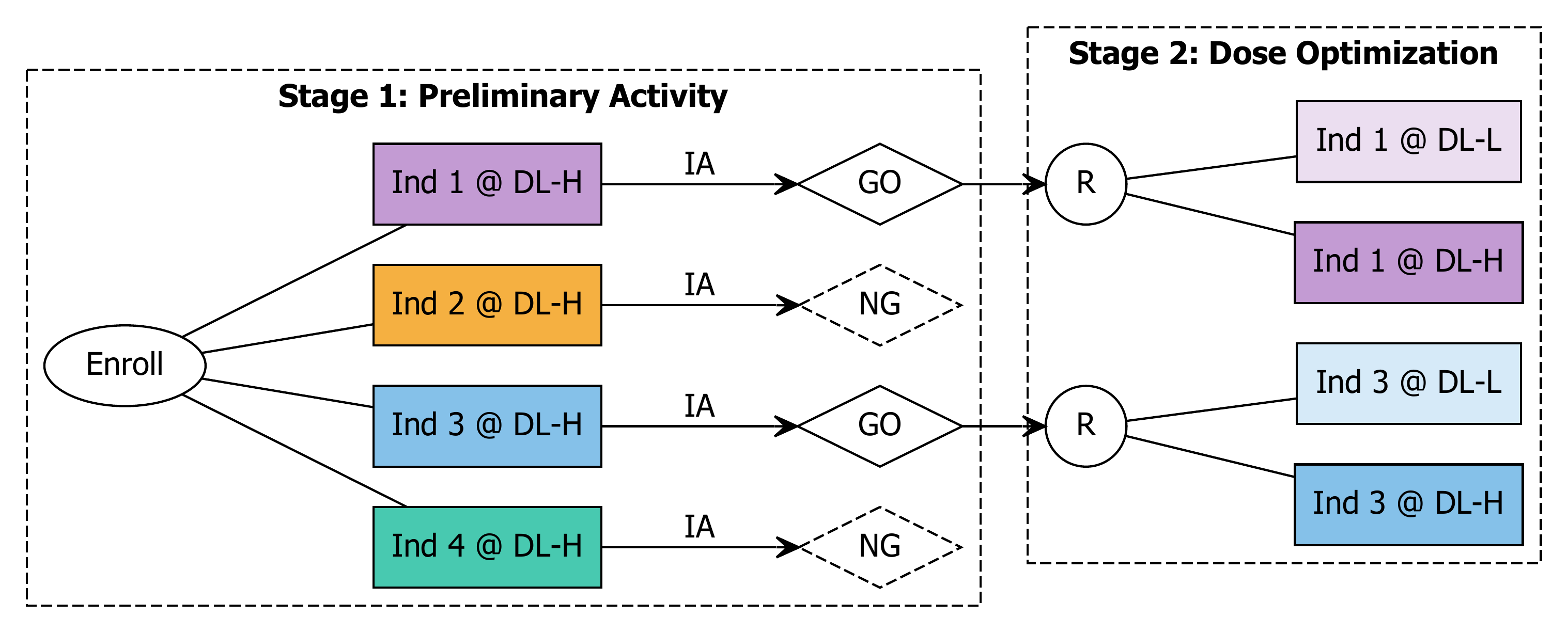}
    \caption{Multi-Arm Two-Stage (MATS) Design Schema. ``NG'' means ``No-Go'' and ``R'' means ``Randomization''.}
    \label{fig:flowchart}
\end{figure}

Figure \ref{fig:flowchart} 
illustrates the overall design schema of MATS using a stylized trial example. A total of four indications ($J=4$) is tested in Stage 1 where participants are treated at the higher dose DL-H. After the interim analysis (IA), two indications (1 and 3) exhibit promising anti-tumor activities and move on to the dose optimization Stage 2 where two dose levels ($I=2$), DL-H and DL-L, are randomized within each of the two indications. The remaining two indications (2 and 4) are stopped due to lack of efficacy (NG: No-Go). At the end of Stage 2, either DL-H or DL-L or none of the two doses will be selected as the optimal dose for each indication. 

\subsection{Decision Criteria}

In this section, we elaborate on the decision criteria and the associated tuning parameters in MATS as multiple decisions at the end of both stages involving PoC or DO will be made. Denote $\tau_{1j}$ as the target (log-odds) difference we wish to distinguish between the response rate of the new investigational drug and the reference response rate (e.g., standard of care). For simplicity, hereinafter we will drop the term  ``log-odds'' unless dropping it would cause a confusion. Denoting Stage 1 data by $\text{data}_1$, at the end of Stage 1 if
\begin{equation}
    \text{P}(\eta_j \geq\tau_{1j}|\text{data}_1)>s_1, \label{eq:s1}
\end{equation}
then the investigational drug is considered to be efficacious for indication $j$ and that indication will continue into Stage 2 (GO); otherwise, no further enrollment to indication $j$ and it is stopped for futility (NG). The decision made at this juncture is denoted by $D_{j1}\in \{0,1\}$, corresponding to the GO and NG decisions, respectively. Here, $s_1$ is a tuning parameter, which can be determined through simulation studies to generate desirable  operating characteristics. To determine $\tau_{1j}$, we first define the \textit{target} response rate $p_{1j}^*$, which is the minimum response rate for which  the investigational drug is considered efficacious. With the specifications of $p_{0j}$ and $p_{1j}^*$, we  define $\tau_{1j}$ as
\begin{equation*}
    \tau_{1j}=\text{logit}\left(\frac{p_{1j}^*+p_{0j}}{2}\right)-\text{logit}(p_{0j}).
\end{equation*}
Alternative definitions for $\tau_{1j}$ such as using $\text{logit}(p_{1j}^*)$ in lieu of $\text{logit}\left(\frac{p_{1j}^*+p_{0j}}{2}\right)$ is possible. 

At the end of Stage 2 (i.e., end of trial), a final analysis is performed. Denoting data from both Stages I and II by $\text{data}_{12}$, the higher dose DL-H is declared efficacious in indication $j$ if 
\begin{equation}
    \text{P}(\eta_j \geq\tau_{1j}|\text{data}_{12})>s_2.\label{eq:s2}
\end{equation}
The lower dose DL-L is declared efficacious in indication $j$ if 
\begin{equation}
    \text{P}(\eta_j-\gamma_j \geq\tau_{1j}|\text{data}_{12})>t_2 .\label{eq:t2}
\end{equation}
And the higher dose DL-H is declared optimal and more efficacious than the lower dose DL-L in indication $j$ if 
\begin{equation}
    \text{P}(\gamma_j \geq\tau_2|\text{data}_{12})>w_2. \label{eq:w2}
\end{equation}

The four different decision criteria are summarized in Table \ref{tab:dec_cri} below.

\begin{table}[htbp]
\centering
\caption{Decision criteria in the MATS design. The ``dose optimization'' in the last row means to test whether efficacy of the higher dose is substantially better than the lower dose. }
\label{tab:dec_cri}
\begin{tabular}{|l|l|l|} 
\hline
Decision criteria & Description & Indicator \\ 
\hline
$\text{P}(\eta_j \geq\tau_{1j}|\text{data}_1)>s_1$ & Determine Go/NG in Stage 1 & $D_{j1}$\\ 
\hline
$\text{P}(\eta_j \geq\tau_{1j}|\text{data}_{12})>s_2$ & Determine PoC of the high dose in Stage 2 & $\mbox{POC-H}_j$ \\ 
\hline
$\text{P}(\eta_j-\gamma_j \geq\tau_{1j}|\text{data}_{12})>t_2$ & Determine  PoC of the low dose in Stage 2 & $\mbox{POC-L}_j$\\
\hline
$\text{P}(\gamma_j \geq\tau_2|\text{data}_{12})>w_2$ & Determine dose optimization in Stage 2 & $\mbox{DO}_j$ \\
\hline
\end{tabular}
\end{table}

For the three decision criteria in Stage 2, $s_2, t_2$ and $w_2$ are tuning parameters, and $\tau_2$ denotes the target difference we wish to distinguish between the response rates of the two dose levels. The value of $\tau_2$ depends on the specific values of $p_{1j}$ and $p_{2j}$, which will be discussed later in Section \ref{sec:tau2}.

Let $\mathbb{I}\{\}$ be an indicator function and let $D_{j1} = \mathbb{I}\{\mbox{Equation } \eqref{eq:s1} \mbox{ is true}\}$ be the GO/NG decision for Stage 1. Also, further define $\mbox{POC-H}_{j} = \mathbb{I}\{\mbox{Equation } \eqref{eq:s2} \mbox{ is true}\}$, $\mbox{POC-L}_{j} = \mathbb{I}\{\mbox{Equation } \eqref{eq:t2} \mbox{ is true}\}$, and $\mbox{DO}_{j} = \mathbb{I}\{\mbox{Equation } \eqref{eq:w2} \mbox{ is true}\}$ be three indicators of conditions \eqref{eq:s2}, \eqref{eq:t2}, and \eqref{eq:w2}, respectively. Using $\mbox{POC-H}_{j}$, $\mbox{POC-L}_{j}$, and $\mbox{DO}_{j}$, for each indication $j$ we estimate the PoC for the high and low doses, as well as the superiority of the high dose over the low dose, i.e., dose optimization. To this end, let $D_{j2}\in \{0,1,2\}$ denote the decision to be made at the end of stage 2 for indication $j$, where  $D_{j2}=1$ means selecting the high dose DL-H as the optimal one, $D_{j2}=2$ means selecting the low dose DL-L as the optimal dose, and $D_{j2}=0$ means that neither dose is selected. Then Algorithm \ref{alg:mats} below presents the MATS design with the focus on the decisions $D_{j1}$ and $D_{j2}$. 

\begin{breakablealgorithm}
\caption{The MATS Trial Design}
\label{alg:mats}
\begin{algorithmic}
    \STATE  Conduct Stage 1 of the trial and record $\text{data}_1 = \{y_{1j1}, n_{1j1}\}$. 
    \STATE  Sample $\eta_j$ from posterior distribution via MCMC. 
    \STATE  Compute $D_{j1} = \mathbb{I}\left(\text{P}(\eta_j \geq\tau_{1j}|\text{data}_1)>s_1 \right)$. 
    \IF{$D_{j1}=1$}
    \STATE  Conduct Stage 2 trials and record data $\{y_{ij2}, n_{ij2}\}$, denote $\text{data}_{12}=\{\text{data}_1, y_{ij2}, n_{ij2}\}$. 
    \STATE  Sample $\eta_j$ and $\gamma_j$ from posterior distribution via MCMC. 
    \STATE  Compute $\mbox{POC-H}_j$, $\mbox{POC-L}_j$, and $\mbox{DO}_j$,
    \STATE  Make Stage 2 decision $D_{j2}$ as follows: 
    \IF{$\mbox{POC-H}_j=\mbox{POC-L}_j=0$,}
    \STATE $D_{j2}=0$ and select no dose;
    \ELSIF{$\mbox{POC-H}_j=\mbox{POC-L}_j=\mbox{DO}_j=1$}
    \STATE $D_{j2}=1$ and select the high dose DL-H;
    \ELSIF{$\mbox{POC-H}_j=1$ and $\mbox{POC-L}_j=0$}
    \STATE $D_{j2}=1$ and select the high dose DL-H;
    \ELSE
    \STATE $D_{j2}=2$ and select the low dose DL-L.
    \ENDIF
    \ELSE
    \STATE Terminate indication $j$ at Stage 1. 
    \ENDIF
\end{algorithmic}
\end{breakablealgorithm}

\subsection{Determination of $\tau_2$}\label{sec:tau2}

\begin{figure}[htbp]
    \centering
    \includegraphics[width=0.8\textwidth]{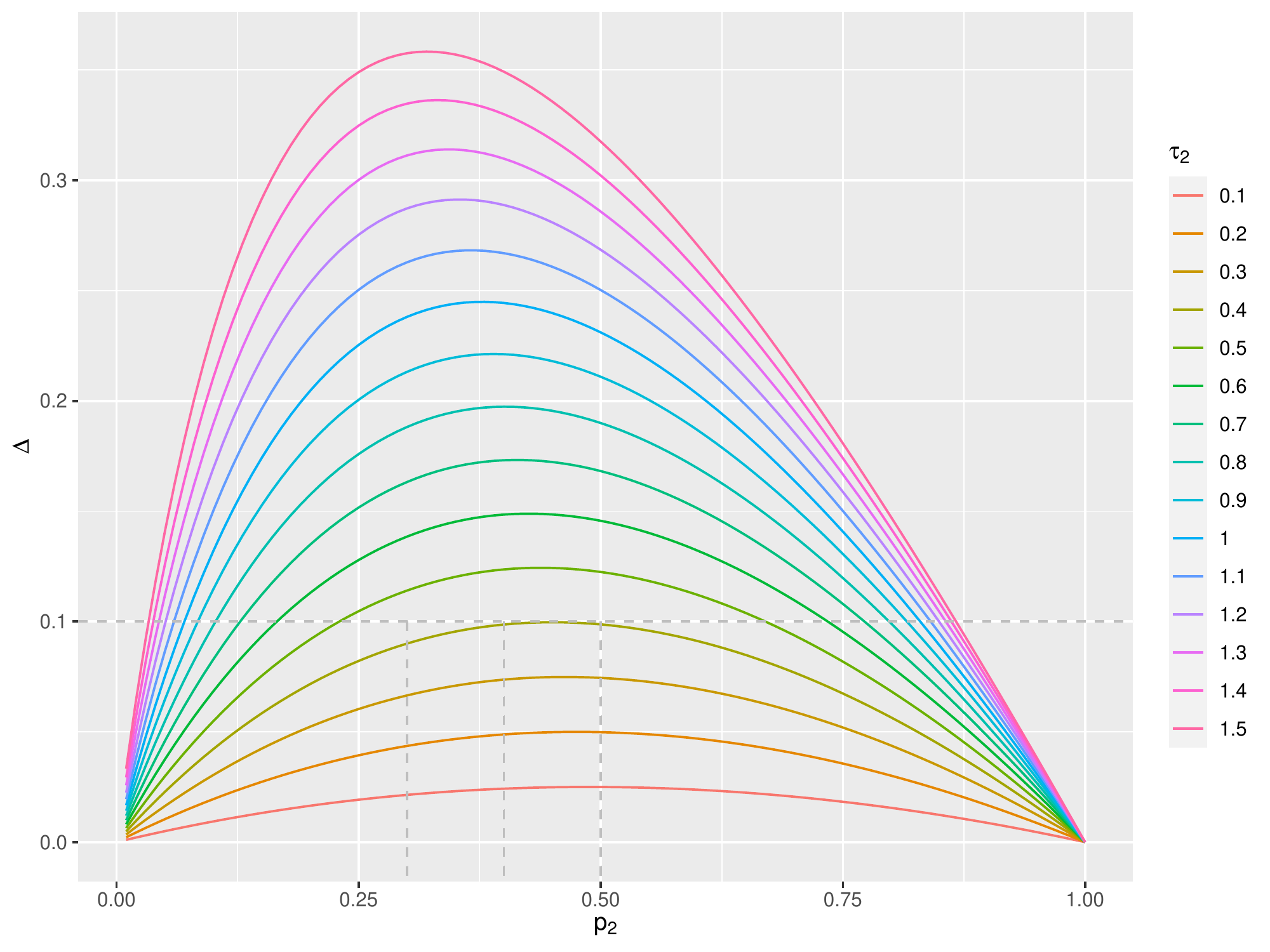}
    \caption{Relation between $p_2$ and the difference value corresponding to the given $\tau_2$}
    \label{fig:tuning}
\end{figure}

In Table \ref{tab:dec_cri}, the value for the tuning parameter $\tau_2$ needs to be determined. 
As mentioned in Algorithm \ref{alg:mats}, $\tau_2$ is used in making the dose-optimization decision in  Stage 2, i.e., $\mbox{DO}_{j}=\mathbb{I}\{\text{P}(\gamma_j \geq\tau_2|\text{data}_2)>w_2\}$. Thus, it serves to determine whether the difference in response rate between the two doses is large enough. However,  $\gamma_j$ denotes the difference on the logit scale and it raises  challenge in calibration of $\gamma_j$ as its relationship to $(p_{1j}-p_{2j})$ is nonlinear. 

To this end, we define $\Delta_j=\text{logit}^{-1}\{\tau_2+\text{logit}(p_{2j})\}-p_{2j}$, which denotes the difference between $p_{1j}$ and $p_{2j}$ according to our model \eqref{eq:model}. Figure \ref{fig:tuning} shows the relationship between $p_{2j}$ and $\Delta_j$ for a grid values of $\tau_2$'s between 0.1 and 1.5. Based on this figure, we propose a procedure to determine $\tau_2$ for a practical trial which requires investigator to determine the minimum difference  between the high and low doses and the possible response rate(s) for the lower dose. 
For example, if we want to distinguish a difference between $p_{1j}$ and $p_{2j}$ greater than or equal to 0.1, we can draw a horizontal line corresponding to $\Delta_j=0.1$ on Figure \ref{fig:tuning}. Assuming the possible response rates for the lower dose are 0.3, 0.4 and 0.5, we then draw vertical lines of $p_{2j}=0.3$, $p_{2j}=0.4$ and $p_{2j}=0.5$. The proposed procedure is to find the maximum $\tau_2$ value such that the intersection points of the corresponding curve with the vertical lines are all below the horizontal line. This turns out to be the curve corresponding to $\tau_2 = 0.4$ in Figure \ref{fig:tuning}. The procedure provides a largest value of $\tau_2$ among the grid such that a true response rate of the high dose that is 0.1 higher than the low dose would be considered superior (i.e., $\gamma_j \ge \tau_2$ would be true), when the low dose response rate is around 0.3, 0.4, or 0.5. In other words, the specified $\tau_2$ value would be appropriate for making $\mbox{DO}_j$ decision if investigators look for a difference of 0.1 in the response rate between the high and low doses.  

\section{Simulation Setup}\label{sec:sim_setup}

\subsection{Scenarios}

We consider a total of eight scenarios in which we specify the true response rate for each dose/indication combination. See Table \ref{tab:sce} for the list of the scenarios. The name of each scenario has a specific meaning: ``GN'' refers to ``global null'', which means that neither dose works in any indication. ``GA-NS'' refers to ``global alternative without superiority'', which means that both doses work across all indications and they have the same efficacy. ``GA-S'' refers to ``global alternative with superiority'', which means that both doses work across all indications but DL-H is more efficacious than DL-L. ``Pick-H-All'' refers to ``pick the higher dose in all indications'', which means that only DL-H works across all indications (while DL-L does not). ``Pick-H-Partial'' refers to ``pick the higher dose in some indications'', which means that only DL-H works in some indications (while DL-L does not). ``Pick-L-Partial'' refers to ``pick the lower dose in some indications'', which means that both doses work in some indications where they have the same efficacy. ``Mixed'' is a mixture of the previous two scenarios. ``Intermediate'' is a variation of ``Pick-H-Partial'', where DL-L is slightly better than placebo for some indications, but not to the extent that it has the same efficacy as DL-H. 

\begin{table}[htbp]
\centering
\caption{True response rates of two doses in four indications for all the scenarios. Dose levels 1 and 2 denote the high dose DL-H and low dose DL-L, respectively. The acronyms for the scenarios are explained in the main text in the Simulation Setup section. }
\label{tab:sce}
\begin{tabular}{c|c|cccc} 
\hline\hline
Scenario & Dose level & Indication1 & Indication2 & Indication3 & Indication4  \\ 
\hline
\multirow{2}{*}{GN} & 1 & 0.1 & 0.2 & 0.1 & 0.2 \\
 & 2 & 0.1 & 0.2 & 0.1 & 0.2 \\ 
\hline
\multirow{2}{*}{GA-NS} & 1 & 0.4 & 0.5 & 0.4 & 0.5 \\
 & 2 & 0.4 & 0.5 & 0.4 & 0.5 \\ 
\hline
\multirow{2}{*}{GA-S} & 1 & 0.5 & 0.6 & 0.5 & 0.6 \\
 & 2 & 0.4 & 0.5 & 0.4 & 0.5 \\ 
\hline
\multirow{2}{*}{Pick-H-All} & 1 & 0.4 & 0.5 & 0.4 & 0.5 \\
 & 2 & 0.1 & 0.2 & 0.1 & 0.2 \\ 
\hline
\multirow{2}{*}{Pick-H-Partial} & 1 & 0.4 & 0.5 & 0.1 & 0.2 \\
 & 2 & 0.1 & 0.2 & 0.1 & 0.2 \\
\hline
\multirow{2}{*}{Pick-L-Partial} & 1 & 0.4 & 0.2 & 0.1 & 0.5 \\
 & 2 & 0.4 & 0.2 & 0.1 & 0.5 \\
\hline
\multirow{2}{*}{Mixed} & 1 & 0.4 & 0.2 & 0.1 & 0.5 \\
 & 2 & 0.4 & 0.2 & 0.1 & 0.2 \\
\hline
\multirow{2}{*}{Intermediate} & 1 & 0.4 & 0.2 & 0.1 & 0.5 \\
 & 2 & 0.3 & 0.2 & 0.1 & 0.4 \\
\hline\hline
\end{tabular}
\end{table}

\section{Simulation Results}\label{sec:sim_results}

\begin{figure}[htbp]
    \centering
    \subfigure[$p_{1j}$]{\includegraphics[width=0.49\textwidth]{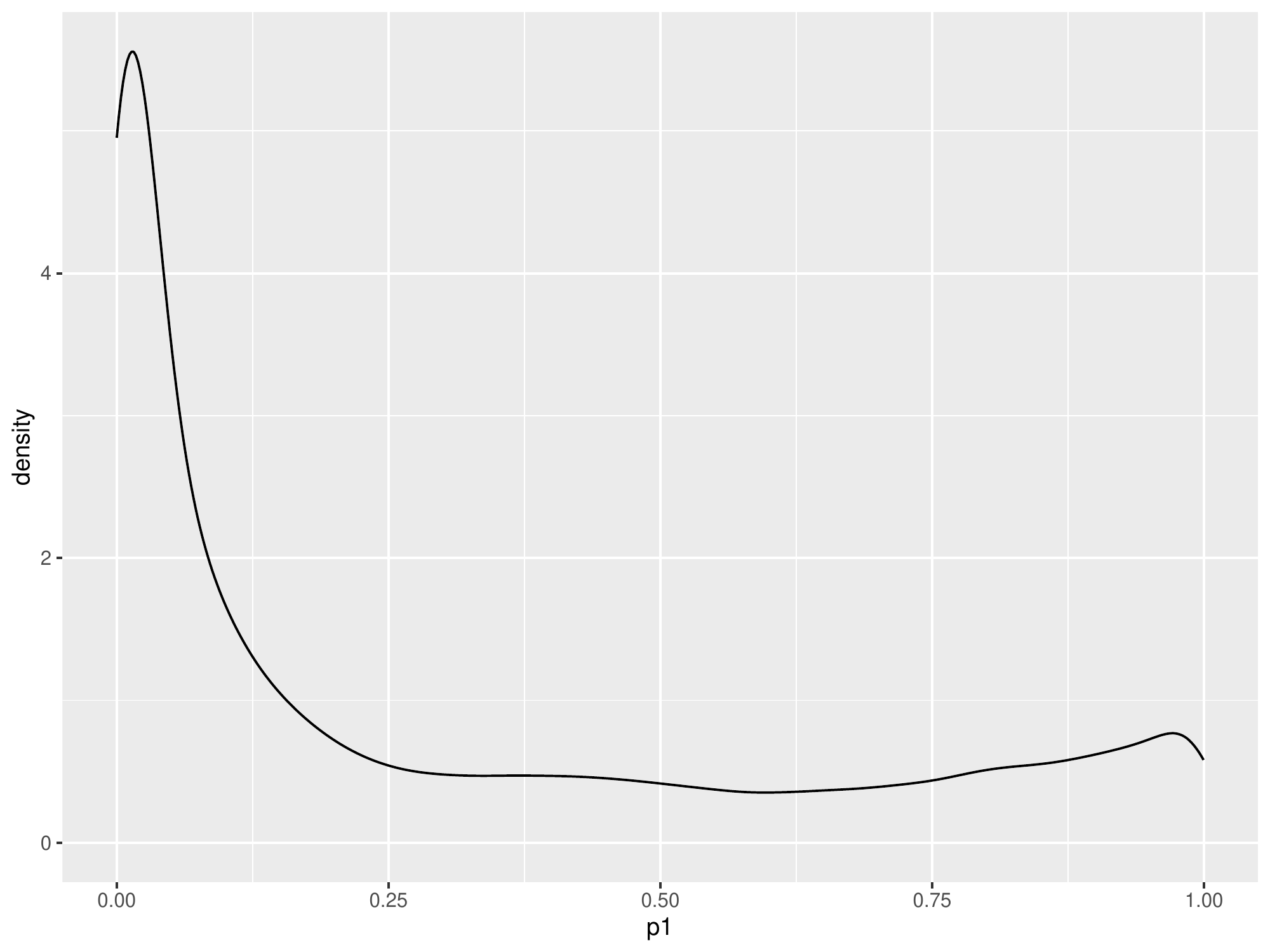}}
    \subfigure[$p_{2j}$]{\includegraphics[width=0.49\textwidth]{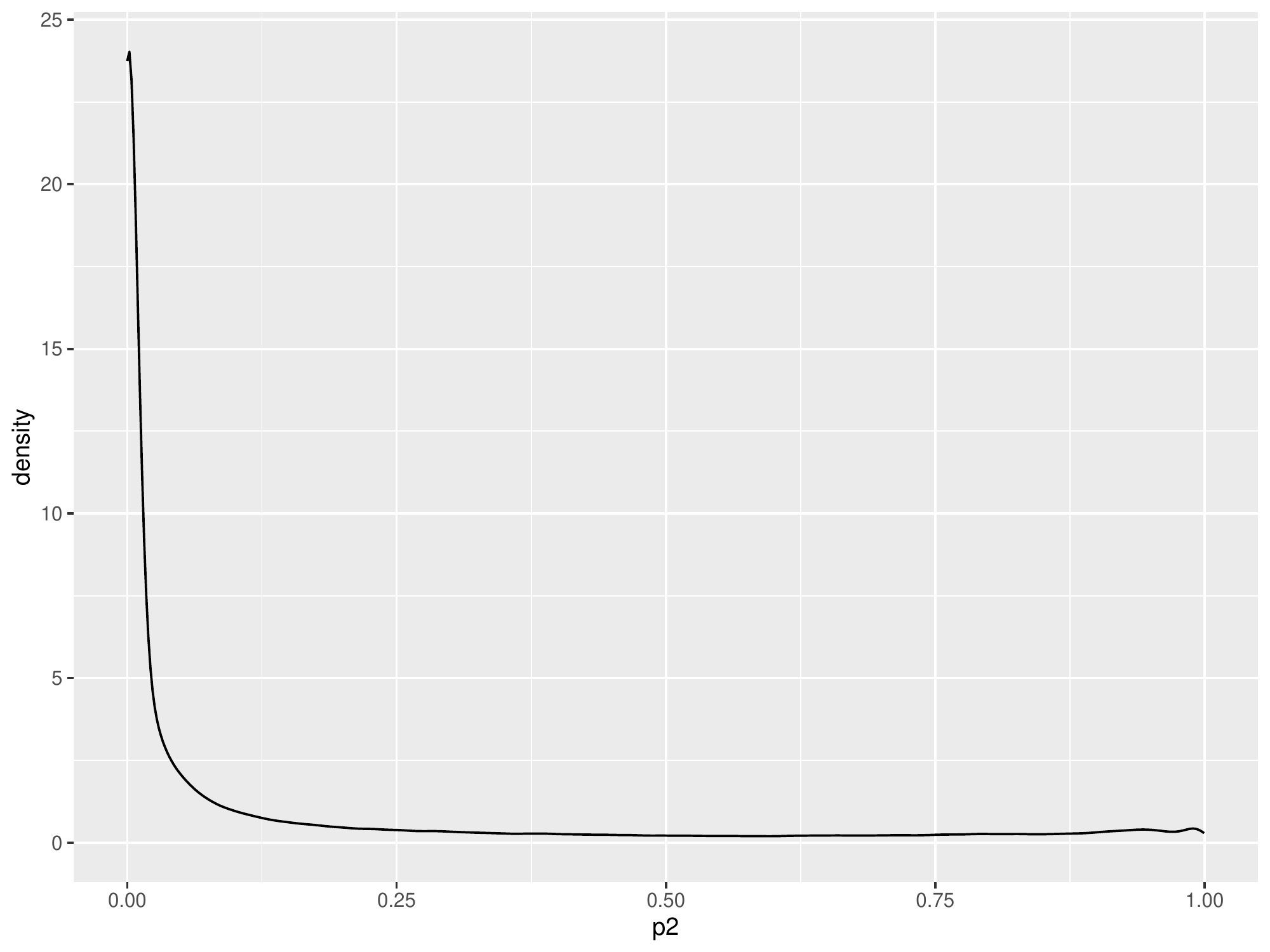}}
    \caption{Prior distributions of $p_{1j}$ and $p_{2j}$ based on the hyperparameter values in the simulation.}
    \label{fig:p1p2prior}
\end{figure}

In our simulation, we set $s_1=s_2=t_2=w_2=0.5$ and $\mu_{\eta_0}=0$, $\sigma_{\eta_0}^2=1$, $\alpha_{\eta}=10$, $\beta_{\eta}=1$, $\mu_{\gamma_0}=0$, $\sigma_{\gamma_0}^2=1$, $\alpha_{\gamma}=2$, $\beta_{\gamma}=1$. These parameters are calibrated to induce vague priors. For example, Figure \ref{fig:p1p2prior} shows the prior distributions of $p_{1j}$ and $p_{2j}$ based on the above hyperparameter values. 

We set the sample size for each dose/indication at each stage to be 20. Tables \ref{tab:perf_met} and \ref{tab:by_ind_error} summarize the operating characteristics of the proposed MATS design. 

\begin{table}[htbp]
\centering
\caption{Operating characteristics of the MATS design for  eight  scenarios with sample size 20 for each dose/indication at each stage. The metrics ``Perfect'', ``PoC'', and ``DO'' refer to the dose selection at each scenario, which is summarized in Table \ref{tab:not} in Appendix A. Type I and Type II error rates are family-wise rates after accounting for multiplicity. }
\label{tab:perf_met}
\begin{tabular}{c|c|c|c|c|c|c} 
\hline\hline
No. & Scenario & \multicolumn{2}{c|}{Stage 1} & \multicolumn{3}{c}{Stage 2} \\ 
\hline
\multicolumn{1}{c}{} & & Type I error rate & Type II error rate & \multicolumn{3}{c}{Type I error rate} \\ 
\hline
1 & GN & 0.124 &  NA & \multicolumn{3}{c}{0.009} \\ 
\hline
\multicolumn{1}{c}{} & \multicolumn{1}{c}{} & \multicolumn{1}{c}{} & & Perfect & PoC & DO \\ 
\hline
2 & GA-NS & NA & 0.324 & 0.380 & 1 & 0.986 \\ 
\hline
3 & GA-S & NA & 0.06 & 0.056 & 1 & 0.689 \\ 
\hline
4 & Pick-H-All & NA & 0.337 & 0.586 & 1 & 1 \\ 
\hline
5 & Pick-H-Partial & 0.062 & 0.214 & 0.717 & 0.980 & 0.980 \\ 
\hline
6 & Pick-L-Partial & 0.072 & 0.206 & 0.507 & 0.993 & 0.887 \\ 
\hline
7 & Mixed & 0.075 & 0.229 & 0.506 & 0.988 & 0.932 \\ 
\hline
8 & Intermediate & 0.072 & 0.225 & 0.238 & 0.673 & 0.673 \\
\hline\hline
\end{tabular}
\end{table}

On the right side of Table \ref{tab:perf_met}, there are three different performance metrics in Stage 2 for scenarios 2 to 8: ``Perfect'' means the correct optimal dose is selected in each indication. ``PoC'' (for ``proof-of-concept'') means the efficacious dose is selected in at least one indication. ``DO'' (for ``dose optimization'') means the correct optimal dose is selected in at least one indication. The same notations are used throughout the article. In Appendix A, the exact definitions of these metrics are provided for reach scenario. Essentially, all three metrics take values between 0 and 1, and the larger value the better performance. They act similar to ``power''. Overall, the MATS design performs well in all the scenarios with reasonable control on the Type I/II error rates and desirable large values for the three metrics. This indicates that  the MATS design is able to weed out nonpromising doses or indications and select the promising ones with high probability. The values for ``Perfect'' metrics sometimes drop very low. This is mainly due to the fact that ``Perfect'' requires correct decisions in all doses (``DO'') and all indications (``PoC'') and the small sample size of early-phase trial (20 per arm in our simulation) makes it difficult to be perfect in all the decisions. It is also noted that this performance metric is highly dependent on the scenarios assumed. Take ``Pick-H-Partial'' (perfect rate = 0.717) for example: it is relatively easier to make a ``perfect'' decision because ``PoC'' of early futility for indications 3 and 4 are an easy call since neither dose of these two indications is efficacious, while ``DO'' for indications 1 and 2 is also not difficult given one dose is efficacious while the other is not. Nevertheless, the ``PoC'' and ``DO'' values are more relevant to early-phase trials here, and the values in Table \ref{tab:perf_met} are all quite reasonable. 

Table \ref{tab:by_ind_error} presents the Type I/II error rates by indication. The values are all quite small for early-phase trials. 


\begin{table}[htbp]
\centering
\caption{Operating characteristics of the MATS design for eight scenarios with sample size 20 for each dose/indication at each stage. Reported are by-indication Stage 1 Type I or II error rates for the MATS with sample size 20 at each dose/indication at each stage. Italic fonts are Type I error rates, and normal fonts are Type II error rates.}
\label{tab:by_ind_error}
\begin{tabular}{c|c|ccccc}
\hline\hline
No. & Scenario & Indication1 & Indication2 & Indication3 & Indication4 \\
\hline
1 & GN & {\it 0.032} & {\it 0.033} & {\it 0.041} & {\it 0.026} \\
\hline
2 & GA-NS & 0.045 & 0.150 & 0.052 & 0.127 \\
\hline
3 & GA-S & 0.007 & 0.028 & 0.006 & 0.021 \\
\hline
4 & Pick-H-All & 0.064 & 0.150 & 0.053 & 0.125 \\
\hline
5 & Pick-H-Partial & 0.056 & 0.168 & {\it 0.036} & {\it 0.026} \\
\hline
6 & Pick-L-Partial & 0.056 & {\it 0.033} & {\it 0.040} & 0.157 \\
\hline
7 & Mixed & 0.052 & {\it 0.027} & {\it 0.050} & 0.185 \\
\hline
8 & Intermediate & 0.060 & {\it 0.029} & {\it 0.045} & 0.174 \\
\hline\hline
\end{tabular}
\end{table}

\begin{figure}[htbp]
    \centering
    \includegraphics[width=0.85\textwidth]{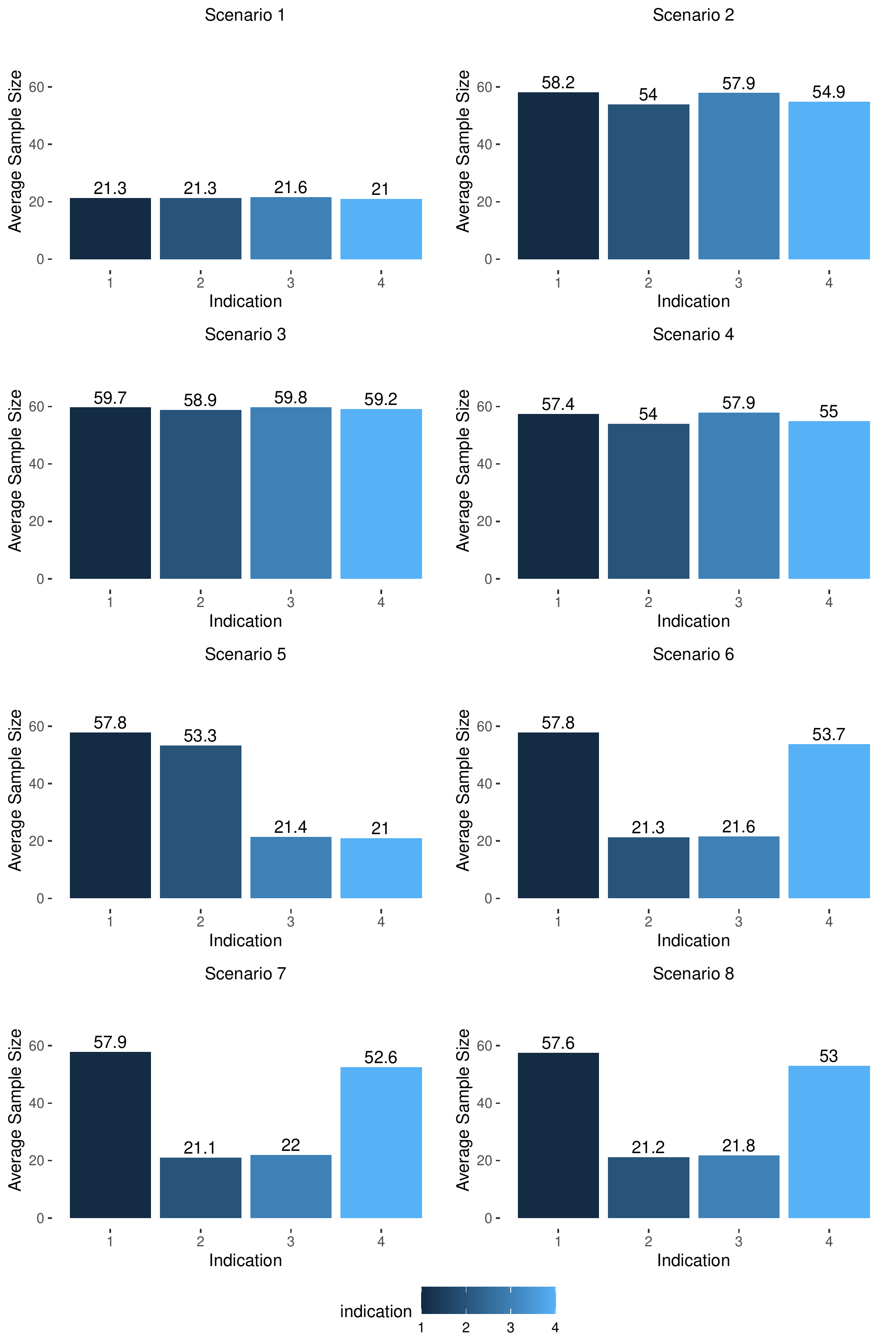}
    \caption{Average sample size of the MATS design in the simulation.}
    \label{fig:ss}
\end{figure}

\begin{figure}[htbp]
    \centering
    \includegraphics[width=0.85\textwidth]{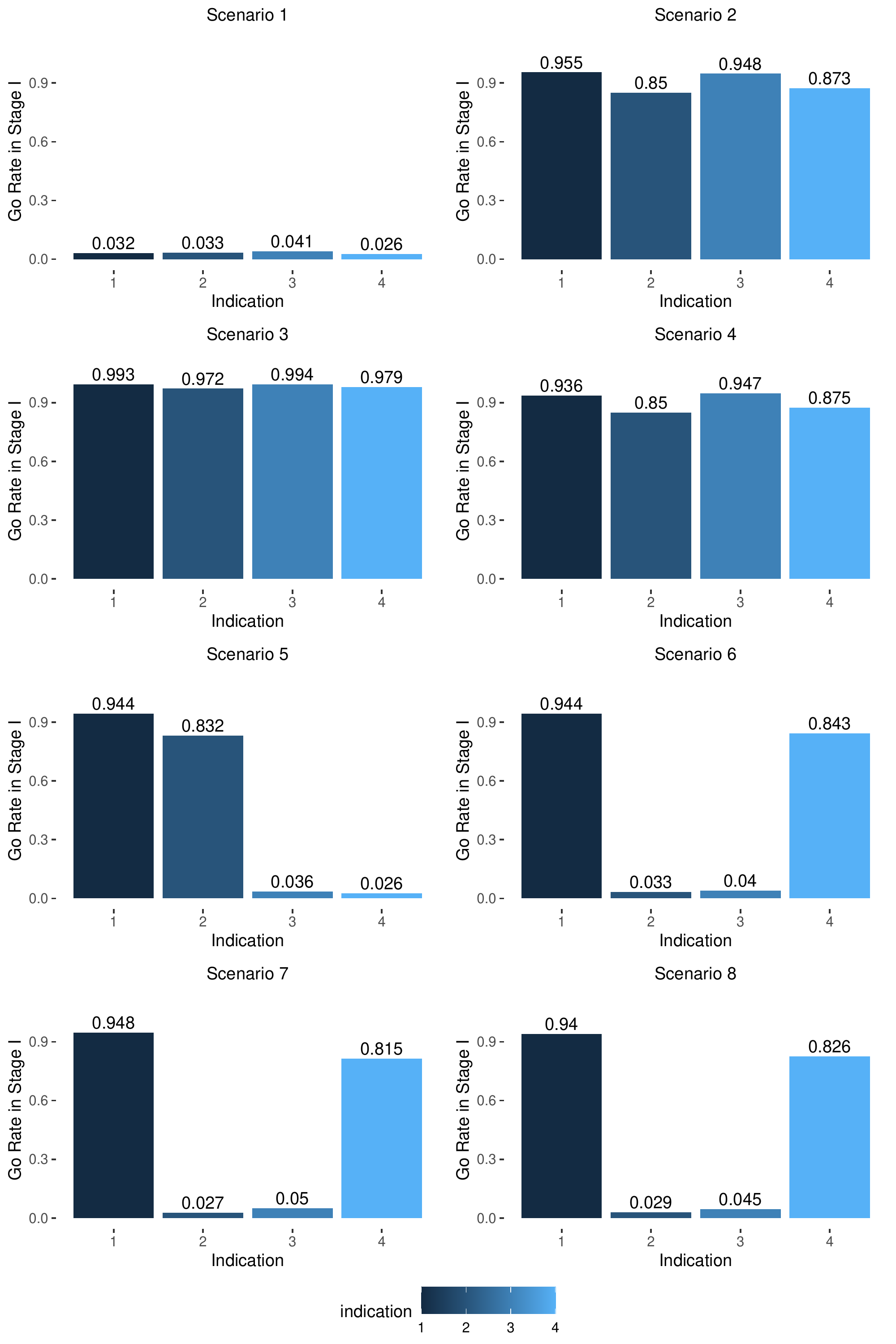}
    \caption{Average GO rates in Stage 1 of the MATS design in the simulation.}
    \label{fig:GO}
\end{figure}

Figures \ref{fig:ss} and \ref{fig:GO} display the average sample size and GO rate in Stage 1 of the MATS design in the simulation for each indication. Recall that 20 is maximum sample size for one dose in one indication at a stage. Therefore, for a high dose and indication that are not promising, they should not proceed to Stage 2 and the arm should be terminated after Stage 1. Otherwise, they should be further tested in Stage 2 with an additional 40 patients randomized at the high and low doses. In other words, a nonpromising arm should exhibit a sample size around 20 and a promising one around 60. We can see this is the case in Figure \ref{fig:ss}. Figure \ref{fig:GO} confirms the GO decision rate for different scenarios. They all look reasonable.

Additional sensitivity simulations are presented in Appendix B, in which we vary the sample size for each dose/indication. Results show that the MATS design are quite robust to small changes of sample sizes.

\section{Discussion}\label{sec:discussion}

The MATS design provides a viable solution to  dose optimization, post dose escalation, when efficacy is  the primary  interest for decision-making. At this stage, endpoints like safety, PK/PD, and exposure, etc., have been investigated in a dose-escalation stage and some doses (such as MTD or MTD-1) appear to be relatively comparable and worth further exploration. One desirable feature of the MATS design is the clever choice of only expanding the high dose DL-H in Stage 1, as opposed to expanding both doses. This choice applies to the situations where the high dose is believed to be more efficacious than the low dose. Therefore, if the high dose is not promising based on Stage 1 data, there is no need to further investigate the low dose in Stage 2. Apparently, for therapeutics where efficacy is not assumed to be monotone across doses, the MATS design will not be appropriate. In order to determine two doses for the MATS design, investigators should have obtained preliminary safety and efficacy signals across multiple ascending candidate doses based on results from the earlier dose-escalation study, from which conventional dose-escalation algorithm's recommendation of the MTD based on incidence of DLTs may be used to determine a candidate for DL-H (or select DL-H as one dose level lower than the MTD if the totality of data support). An intermediate dose level with initial efficacy or a dose level that first shows ``proof-of-mechanism'' may be considered as a candidate for DL-L. Additional evidence from PK/PD to characterize the dose-/exposure-response curve, albeit preliminary, should also be taken into consideration.

The proposed MATS design is in accordance with a few key considerations outlined in the FDA Guidance for Industry on dose optimization \citep{fda2023optimizing}: 

\begin{enumerate}
    \item The Guidance states that ``\textit{Clinical trials should enroll an appropriately \textbf{broad population} to allow assessment of the dosage(s) across relevant subpopulations.}'' The MATS design simultaneously considers multiple indications in a randomized setting, allowing for the exploration of multiple potentially promising subpopulations at the early stage while keeping the expected total sample size under a reasonable range. 
    \item The Guidance states that ``\textit{Sponsors, \dots, should plan their development programs such that identification of the optimal dosage(s) can occur prior to or \textbf{concurrently with} the establishment of the drug’s safety and effectiveness.}'' The MATS design seamlessly connects Stage 1 for preliminary activity and safety evaluations with the dose optimization stage that follows, substantially boosting the efficiency of development and expediting the timeline. 
    \item The Guidance states that ``\textit{An \textbf{adaptive design} to stop enrollment of participants to one or more dosage arms of a clinical trial following an interim assessment of efficacy and/or safety could be considered.}'' The MATS design is adaptive in nature so that dose optimization involving more participants will only be triggered upon observing preliminary activity and acceptable safety profile at the interim analysis (end of Stage 1). 
\end{enumerate}

The  MATS design is flexible in striking a balance between the desired type I and type II error rates by appropriate calibration of the thresholds. Because such dose optimization studies are not intended for registration (i.e., not designed to establish superiority of the drug to a comparator), depending on factors such as sponsor’s pipeline prioritization and foreseeable competitive landscape, more forgiving goals for controlling statistical errors may be set. On one hand, the turning parameters can be set to ensure a relatively high probability of advancing a promising drug from Stage 1 to Stage 2 (i.e., lower the type II error rates under the GA scenario) at the expense of slightly inflated false positive rates (reflected by Stage 1 by-indication type I error rates under the GN scenario). On the other hand, the tuning parameters can also bet set to ensure a relatively low probability of advancing a non-promising drug from Stage 1 to Stage 2 (reflected by Stage 1 by-indication type I error rates controlled below 10\% under the GN scenario) at the expense of slightly inflated false negative rates (reflected by Stage 1 by-indication type II error rates under the GA scenario). We consider this flexibility as a very appealing feature by the MATS design.

The Stage 1 in MATS for preliminary activity is necessary because according to \citet{wong2019estimation}, the probability of success rate in oncology is only 57.6\% from phase 1 to phase 2, and 32.7\% from phase 2 to phase 3, both of which are the lowest compared to other therapeutic groups. Without a preliminary activity signal of the drug, dose optimization based on clinical efficacy is infeasible and may only expose more than necessary participants to inefficacious treatment regimen. Having a small expansion cohort before dose optimization is also one of the feedback we received from FDA (although this recommendation may not generalize to every study as the dose optimization design should be considered on a case-by-case basis). In this article, we assume that the overall response rate (ORR) is a reliable measure of a drug's efficacy and thus chosen as the endpoint of interest at both stages. Alternative endpoint(s) should be considered if ORR turns out to be a poor predictor of the ultimate clinical outcome, but the overall two-stage adaptive design framework remains the same.

While the sample size in Stage 1 is minimal at 20 and it may be subject to change depending on the specific considerations of a study; e.g., a small cohort size at the presumptive RP2D is suggested in the \textit{FDA Guidance for Industry on Acute Myeloid Leukemia} \citep{fda2022acute}. For example, to prevent exposing unnecessary too many participants to the higher dose which may be more toxic than expected, a continuous safety monitoring rule can be established in Stage 1 \citep{ivanova2005continuous}. Such precautionary measure for safety, although not quantitatively incorporated with an efficacy endpoint via joint modeling, is still indispensable as part of the dose optimization process.

Another advantage of the MATS design is its flexibility of handling different doses for different indications which are quite possible for combination regimens. Different MTDs/MADs may be identified by the end of the dose-escalation phase as the evaluations are typically performed for each combination specific for the indication of interest. Take ABRAXANE\textregistered (paclitaxel) \citep{fda2020abraxane} for example as it has different approved doses among different combinations/indications : the recommended dosage is 260 mg/$\text{m}^2$ as a single-agent for metastatic breast cancer, 100 mg/$\text{m}^2$ when combined with carboplatin for non-small cell lung cancer, and 125 mg/$\text{m}^2$ when combined with gemcitabine for pancreatic cancer. Instead of running multiple dose optimizations separately for each indication, MATS could naturally address such situation by allowing different doses for different indications (i.e., the ``DL-H'' doses at Stage 1 are not necessarily the same across indications) while enabling information-borrowing across combinations/indications with the assumption of constant log-odds difference between the two dose levels tested in each indication. As a result, MATS is expected to substantially expedite the development timeline to deliver novel agents to patients in urgent needs. 

We would like to emphasize that the MATS design is not intended to provide a ``one-fits-all'' solution to dose optimization: in practice, the study team should still incorporate the totality of data (or some composite endpoint may be defined) into the data package for the dose optimization discussion with regulatory agencies. In addition to the FDA Guidance for Industry on dose optimization, the following Guidance documents are also relevant to this topic and informative for the preparation of the complete data package: \textit{Population Pharmacokinetics} \citep{fda2019population} and \textit{Exposure-Response Relationships - Study Design, Data Analysis, and Regulatory Applications} \citep{fda2003exposure}.   

Based on the MATS design, we could  consider a mixture model. In this way, the strength of borrowing between different indications can be determined by the data. To achieve this, we could use finite mixture model in $\gamma_j$ and modify its prior to

\begin{equation*}
    \gamma_j|x_j=t\sim \text{LogNormal}(\gamma_0, \sigma_{\gamma,t}^2), t=0,1
\end{equation*}

where $x_j\in\{0,1\}$ is a latent variable denoting whether the prior of $\gamma_j$ is informative or not. Specifically, $x_j=0$ refers to an informative prior for $\gamma_j$ while $x_j=1$ refers to a non-informative one. The hyperparameters $\gamma_0$ and $\sigma_{\gamma,t}^2$ are given hyperpriors by

\begin{equation*}
    \gamma_0|\mu_{\gamma_0},\sigma^2_{\gamma_0} \sim \text{N}(\mu_{\gamma_0},\sigma^2_{\gamma_0}) \quad\text{and}\quad \sigma^2_{\gamma,t}|\alpha_{\gamma,t},\beta_{\gamma,t} \sim \text{Inv-Gamma}(\alpha_{\gamma,t},\beta_{\gamma,t}), t=0,1
\end{equation*}

As is shown, to identify the above-mentioned difference in the information contained, different hyperpriors are given to the hyperparameter $\sigma_{\gamma,t}^2$, which is related to the variance of the prior. \jj 

Based on our recent feedback from the FDA regarding dose optimization across multiple oncology projects, potential extensions of the MATS design could be made to ``fit for purpose'': one extension is to ``reverse'' the two stages in MATS by first starting with the randomized dose optimization of two dose levels for each indication considered, followed by an expansion cohort of one dose level upon observing promising efficacy and favorable safety for some or all indications. Compared to MATS, this design variant may result in a larger total sample size, especially when there are only few ``promising'' indications (because the number of participants enrolled to those indications is approximately doubled compared to that in the MATS design). However, it could be advantageous when the lower dose turns out to be optimal, as the MATS design will not evaluate the lower dose until the later Stage 2. Future work is needed to compare MATS with different variants of the design that fit the needs under different scenarios.

To facilitate the implementation of the MATS design in practice, we have developed an R Shiny application that includes trial simulations, data analyses and tuning parameter determinations. It is available at \url{https://matsdesign.shinyapps.io/mats/}. 

\newpage

\nocite{*} 

\bibliographystyle{plainnat}
\bibliography{main}

\newpage

\begin{appendices}

\counterwithin{table}{section} 

\renewcommand{\thetable}{\Alph{section}.\arabic{table}} 
\renewcommand{\thefigure}{\Alph{section}.\arabic{figure}} 

\section{Some Notation}

\begin{table}[htbp]
\centering
\caption{Specific meaning of performance metrics in the different indications}
\label{tab:not}
\scalebox{0.93}{
\begin{tabular}{|c|l|} 
\hline
\textbf{Scenario} & \multicolumn{1}{c|}{\textbf{Performance Metrics}} \\ 
\hline
GN & Type I error rate: \% of selecting ANY dose in Stage 2 for ANY indication \\ 
\hline
GA-NS & \begin{tabular}[c]{@{}l@{}}``Perfect'': \% of selecting DL2 in ALL indications\\ ``PoC'': \% of selecting DL1 OR DL2 in ANY indication\\ ``DO'': \% of selecting DL2 in ANY indication\end{tabular} \\ 
\hline
GA-S & \begin{tabular}[c]{@{}l@{}}``Perfect'': \% of selecting DL1 in ALL indications\\ ``PoC'': \% of selecting DL1 OR DL2 in ANY indication\\ ``DO'': \% of selecting DL1 in ANY indication\end{tabular} \\ 
\hline
Pick-H-All & \begin{tabular}[c]{@{}l@{}}``Perfect'': \% of selecting DL1 in ALL indications\\ ``PoC'': \% of selecting DL1 in ANY indication\\ ``DO'': \% of selecting DL1 in ANY indication\end{tabular} \\ 
\hline
Pick-H-Partial & \begin{tabular}[c]{@{}l@{}}``Perfect'': \% of selecting DL1 in indications 1 AND 2\\ ``PoC'': \% of selecting DL1 in indication 1 OR 2\\ ``DO'': \% of selecting DL1 in indication 1 OR 2\end{tabular} \\ 
\hline
Pick-L-Partial & \begin{tabular}[c]{@{}l@{}}``Perfect'': \% of selecting DL2 in indications 1 AND 4\\ ``PoC'': \% of selecting DL1 OR DL2 in indication 1 OR 4\\ ``DO'': \% of selecting DL2 in indication 1 OR 4\end{tabular} \\ 
\hline
Mixed & \begin{tabular}[c]{@{}l@{}}``Perfect'': \% of selecting DL1 in indication 4 AND selecting DL2 in indication 1\\ ``PoC'': \% of selecting DL1 OR DL2 in indication 1 OR selecting DL1 in indication 4\\ ``DO'': \% of selecting DL1 in indication 4 OR selecting DL2 in indication 1\end{tabular} \\ 
\hline
Intermediate & \begin{tabular}[c]{@{}l@{}}``Perfect'': \% of selecting DL1 in indication 1 AND 4\\ ``PoC'': \% of selecting DL1 in indication 1 OR 4\\ ``DO'': \% of selecting DL1 in indication 1 OR 4\end{tabular} \\
\hline
\end{tabular}
}
\end{table}

\section{Sensitivity Analysis}

\subsection{Add Sample Size}

We can change the sample size from 20 to 30. 

\begin{table}[htbp]
\centering
\caption{Operating characteristics of the MATS design for eight scenarios scenarios with sample size 30 for each dose/indication at each stage. The metrics ``Perfect'', ``PoC'', and ``DO'' refer to the dose selection at each scenario, which is summarized in Table \ref{tab:not} in Appendix A. Type I and Type II error rates are family-wise rates after accounting for multiplicity.}
\begin{tabular}{c|c|c|c|c|c|c} 
\hline\hline
No. & Scenario & \multicolumn{2}{c|}{Stage 1} & \multicolumn{3}{c}{Stage 2} \\ 
\hline
\multicolumn{1}{c}{} & & Type I error rate & Type II error rate & \multicolumn{3}{c}{Type I error rate} \\ 
\hline
1 & GN & 0.052 & NA & \multicolumn{3}{c}{0} \\ 
\hline
\multicolumn{1}{c}{} & \multicolumn{1}{c}{} & \multicolumn{1}{c}{} & & Perfect & PoC & DO \\ 
\hline
2 & GA-NS & NA & 0.231 & 0.568 & 1 & 0.998 \\ 
\hline
3 & GA-S & NA & 0.014 & 0.057 & 1 & 0.666 \\ 
\hline
4 & Pick-H-All & NA & 0.221 & 0.741 & 1 & 1 \\ 
\hline
5 & Pick-H-Partial & 0.037 & 0.124 & 0.851 & 0.997 & 0.997 \\ 
\hline
6 & Pick-L-Partial & 0.030 & 0.134 & 0.687 & 0.995 & 0.947 \\ 
\hline
7 & Mixed & 0.030 & 0.131 & 0.653 & 0.996 & 0.972 \\ 
\hline
8 & Intermediate & 0.028 & 0.130 & 0.253 & 0.665 & 0.665 \\
\hline\hline
\end{tabular}
\end{table}

\begin{table}[htbp]
\centering
\caption{Operating characteristics of the MATS design for eight scenarios scenarios with sample size 30 for each dose/indication at each stage. Reported are by-indication Stage 1 Type I or II error rates for the MATS with sample size 30 at each dose/indication at each stage. Italic fonts are Type I error rates, and normal fonts are Type II error rates.}
\begin{tabular}{c|c|ccccc}
\hline\hline
No. & Scenario & Indication1 & Indication2 & Indication3 & Indication4 \\
\hline
1 & GN & {\it 0.018} & {\it 0.008} & {\it 0.022} & {\it 0.007} \\
\hline
2 & GA-NS & 0.012 & 0.112 & 0.018 & 0.109 \\
\hline
3 & GA-S & 0.001 & 0.006 & 0 & 0.007 \\
\hline
4 & Pick-H-All & 0.012 & 0.105 & 0.019 & 0.103 \\
\hline
5 & Pick-H-Partial & 0.024 & 0.1 & {\it 0.023} & {\it 0.014} \\
\hline
6 & Pick-L-Partial & 0.023 & {\it 0.009} & {\it 0.021} & 0.116 \\
\hline
7 & Mixed & 0.024 & {\it 0.009} & {\it 0.021} & 0.111 \\
\hline
8 & Intermediate & 0.028 & {\it 0.009} & {\it 0.019} & 0.104 \\
\hline\hline
\end{tabular}
\end{table}

\subsection{Unbalanced Stage 2 Sample Size}

At the beginning of Stage 2, we already have some information about the performance at the high dose from trials in Stage 1, while no information at the low one. Thus, it is reasonable to assign more samples to the low dose than the high dose in Stage 2. 

\begin{table}[htbp]
\centering
\caption{Operating characteristics of the MATS design for eight scenarios scenarios with sample size 20 at stage 1, 10 at stage 2 in high dose and 20 in low dose for each indication. The metrics ``Perfect'', ``PoC'', and ``DO'' refer to the dose selection at each scenario, which is summarized in Table \ref{tab:not} in Appendix A. Type I and Type II error rates are family-wise rates after accounting for multiplicity.}
\begin{tabular}{c|c|c|c|c|c|c} 
\hline\hline
No. & Scenario & \multicolumn{2}{c|}{Stage 1} & \multicolumn{3}{c}{Stage 2} \\ 
\hline
\multicolumn{1}{c}{} & & Type I error rate & Type II error rate & \multicolumn{3}{c}{Type I error rate} \\ 
\hline
1 & GN & 0.12 & NA & \multicolumn{3}{c}{0.025} \\ 
\hline
\multicolumn{1}{c}{} & \multicolumn{1}{c}{} & \multicolumn{1}{c}{} & & Perfect & PoC & DO \\ 
\hline
2 & GA-NS & NA & 0.336 & 0.377 & 1 & 0.989 \\ 
\hline
3 & GA-S & NA & 0.056 & 0.058 & 1 & 0.662 \\ 
\hline
4 & Pick-H-All & NA & 0.332 & 0.578 & 1 & 1 \\ 
\hline
5 & Pick-H-Partial & 0.062 & 0.194 & 0.723 & 0.977 & 0.977 \\ 
\hline
6 & Pick-L-Partial & 0.079 & 0.219 & 0.493 & 0.989 & 0.887 \\ 
\hline
7 & Mixed & 0.071 & 0.216 & 0.502 & 0.99 & 0.937 \\ 
\hline
8 & Intermediate & 0.072 & 0.219 & 0.231 & 0.679 & 0.679 \\
\hline\hline
\end{tabular}
\end{table}

\begin{table}[htbp]
\centering
\caption{Operating characteristics of the MATS design for  eight scenarios scenarios with sample size 20 at stage 1, 10 at stage 2 in high dose and 20 in low dose for each indication. Reported are by-indication Stage 1 Type I or II error rates for the MATS with sample size 20 at stage 1, 10 at stage 2 in high dose and 20 in low dose for each indication. Italic fonts are Type I error rates, and normal fonts are Type II error rates.}
\begin{tabular}{c|c|ccccc}
\hline\hline
No. & Scenario & Indication1 & Indication2 & Indication3 & Indication4 \\
\hline
1 & GN & {\it 0.038} & {\it 0.035} & {\it 0.037} & {\it 0.020} \\
\hline
2 & GA-NS & 0.051 & 0.161 & 0.049 & 0.131 \\
\hline
3 & GA-S & 0.009 & 0.031 & 0.004 & 0.012 \\
\hline
4 & Pick-H-All & 0.045 & 0.142 & 0.054 & 0.133 \\
\hline
5 & Pick-H-Partial & 0.052 & 0.151 & {\it 0.035} & {\it 0.027} \\
\hline
6 & Pick-L-Partial & 0.059 & {\it 0.036} & {\it 0.045} & 0.170 \\
\hline
7 & Mixed & 0.048 & {\it 0.034} & {\it 0.039} & 0.177 \\
\hline
8 & Intermediate & 0.062 & {\it 0.032} & {\it 0.042} & 0.168 \\
\hline\hline
\end{tabular}
\end{table}

\end{appendices}

\end{document}